\documentclass[aps,prc,showpacs,nofootinbib,twocolumn]{revtex4-1}

\usepackage{epsfig}
\usepackage{graphicx,amsmath}
\usepackage{color}
\newcommand\ba{\begin{eqnarray}}
\newcommand\ea{\end{eqnarray}}

\newcommand{\be}{\begin{equation}}
\newcommand{\ee}{\end{equation}}
\newcommand{\bas}{\begin{eqnarray*}}
\newcommand{\eas}{\end{eqnarray*}}
\begin{document}
\vspace*{-1.5cm}
\title{Possible existence of a meson (${\it s}{\bar s}$) S = 0 at M $\approx$ 762~MeV } 
\author{B. Tatischeff}
\email{tati@ipno.in2p3.fr}
\affiliation{CNRS/IN2P3, Institut de Physique Nucl\'eaire, UMR 8608, 
 and Univ. Paris-Sud, 91405 Orsay, France}
 
\author{E. Tomasi-Gustafsson}
\email{egle.tomasi@cea.fr}
\affiliation{IRFU/SPhN, CEA/Saclay, 91191 Gif-sur-Yvette Cedex, France}
\pacs{12.40.Yx, 13.20.Jf, 13.75.Cs, 14.40.Be}
\vspace*{1cm}
\begin{abstract}
The possible existence of a ($ s\bar s$) S = 0 meson at M $\approx$ 762~MeV is discussed through a critical analysis of the existing data. Different experimental results are considered and show the possibility that the presence of such meson is not excluded by the data, but may be hidden by  more excited mesons at nearby masses.
\end{abstract}
\maketitle

\section{Introduction}
The study of the meson spectroscopy remains important, but presently mostly  restricted to mass range larger than 3~GeV, like charmonium or bottomonium. Indeed many measurements are performed exploiting the collider properties.
The calculations concerning charmonium or bottomonium are facilitated since heavy quarks are involved, allowing application of perturbation theories. These studies are stimulated by the experimental observation of isospin 1 mesons, forbidden in simple $q\bar q$ models.
 However open questions remain at lower masses, like, for example, the spin determination of several charmed strange, bottom, or bottom strange mesons. This paper shows that, in addition to new experimental data and theoretical results, it is possible to get information on meson spectroscopy, using simple well known properties, not yet applied for that purpose.

Although the mesons are rather well described in the relativized quark model with chromodynamics
\cite{godfrey}, the authors mentioned the difficulty - they wrote "the problem is especially severe"  - to describe the isoscalar-pseudoscalar mesons. There is no state found in their approach like the one discussed here.

It was noted \cite{amsler3} that "lightest isoscalar pseudoscalar mesons may mix more strongly with excited states or with states of substantial non-$ q\bar q$ content." Different 
${\eta-\eta'}$ mixing angles were advocated, as well as possible gluonic and intrinsic $c\bar c$ components. Here also, the authors wrote that "the situation for the pseudoscalar and scalar mesons is not so clear cut, either theoretically or experimentally." A theoretical solution of the lack of the missing  $s\bar s$ (S = 0) meson mass was suggested  in a work stressing that the "chiral U(1) symmetry is explicitly broken by instantons" \cite{hooft}. 
The present paper is devoted to the study of the experimentally "missing" low mass S = 0, meson ($s\bar s$)  called "M". 

From the experimental point of view, we are faced to the following questions:
\begin{enumerate}
\item if this meson does not exist, is there an explanation to its non observation~?
\item otherwise, what is its mass ?
\item what reaction should be used to allow its possible observation ?
\end{enumerate}
It was pointed out long time ago, that the chromomagnetic interaction used in atomic physics, is able to describe the hyperfine splittings in hadrons. This old, simple model, is no more considered, since overtaken by modern theoretical approaches. It can however be used to answer the second question noted above, namely to give an indication of the possible mass.  Several low meson masses can be reproduced using the spin-spin coupling of quarks, similarly to the hyperfine splitting in electromagnetic interaction. The formula used is:
 \be
 m({\it q}{\bar q}) = m_1  + m_2 + A {\vec S_1}.{\vec S_2}/m_1 m_2,
 \ee
where $S_{1}$ and $S_{2}$ are the spins of both quarks. $m_1$ and $m_2$ are the constituent quark masses. A defines the amount of the difference ($\Delta$E) between the masses of the first S = 1 and the ground S = 0 of the studied meson family. A = $(2*m(u))^{2}$c. 
 Such hyperfine interaction suggests for hadrons, that there is a QCD source in single gluon exchange \cite{close}. 
 
 Table~I shows the masses calculated using the following parameters: \\
- the "u" and "d" (constituent quark) masses are m(u) = 309~MeV,\\
- the constituent strange quark mass m(s) = 480~MeV,\\
- "c" = 159~MeV.
 
 A  line is added, which corresponds to (${\it s}{\bar s}$) S = 0.
 \begin{table}[h]
\vspace*{-0.2cm}
\caption{Low meson masses (in MeV) calculated by using equation (1).}
\label{Table 1}
\vspace{0.mm}
\begin{tabular}{c c c c c c c c}
\hline
meson&$m_{1}$&$m_{2}$&c&S&Calc.&Exp.&$\mid{\Delta M} |$/M \\
\hline
$\pi$&309&309&159&0&141&138&2.2 $10^{-2}$\\
$\rho$&309&309&159&1&777.0&775.5&1.9 $10^{-3}$\\
$\omega$&309&309&159&1&777&782.6&7.2 $10^{-3}$\\
K&309&480&159&0&481.9&481.7&4.2 $10^{-4}$\\
$K^{*}$&309&480&159&1&891.4&892&6.7 $10^{-4}$\\
(${\it s}{\bar s}$)&480&480&159&0&762.3&?&\\
$\phi$&480&480&159&1&1025.9&1019.5&6.3 $10^{-3}$\\
\hline
\end{tabular}
\end{table}
We observe that the masses are well reproduced. The $\phi$ meson is considered as a (${\it s}{\bar s}$) meson with spin S = 1. A  (${\it s}{\bar s}$) meson with spin S = 0, is predicted to exist at M $\simeq$ 762.3. Such meson has never been observed, neither predicted.

Another still rough mass relation, assuming that the masses follow  a: $a +b S(S+1)$  variation, predicts a value for the "M" mass close to the the previous one. We already know several mesons,  having an important (${\it s}{\bar s}$) component, namely the $\phi$ (1020) S = 1, and the $f'_{2}$ (1525) (S = 2). Using this  relation  we get M(S = 0) $\approx$ 767~MeV.

Let us stress that we do not take these results as theoretical predictions, but only as an indication of the mass range where such meson should be looked for.

Along this study the "average masses" given by \cite{pdg} are used.

\section{Possible explanation about its non observation, and study of favoured reactions}

The possible mass of the "M" meson  is too low to allow its decay into two kaons. Its mass is likely lower than $\rho$ or $\omega$ masses, allowing only disintegration modes with pions and (or) gammas.  

This meson should have a orbital momentum $\ell= 0$ between both quarks, therefore S = I = 0, then P = -1, C = +1, and G = +1. Then its quantum numbers $I^{G}(J^{PC})$ are $0^{+}(0^{-+})$ and are the same as those of the $\eta$(548), making him a possible candidate for mixing with the $\eta$(548) instead of the $\eta'$(958). This will be discussed below.

The positive charge conjugation shows directly the possible disintegration modes, since the $\rho$ and $\omega$ mesons, having both negative charge conjugation have different disintegration modes. 
Parity conservation forbids the "M" disintegration into two $\pi^{0}$, or $\pi^{+} \pi^{-}$ by strong or electromagnetic interactions. The $\eta\pi^{0}$ mode is forbidden by isospin conservation. 

The disintegration modes being different from those of $\rho$ or $\omega$ mesons,  we anticipate the existence of a not too small (not hidden) signature into allowed disintegration modes of "M".

The  possible modes for "M" are 2$\gamma$ or three pions (3$\pi^{0}$, or $\pi^{+} \pi^{-} \pi^{0}$). 
The $(\pi^{+} \pi^{-} \gamma)$ mode is not forbidden, but its branching ratio is small 4.6 \% for 
$\eta$(548) and decreases for increasing meson mass. Indeed it is not given for $\eta'$(958) but only included in the resonant $\rho \gamma$ decay mode  \cite{pdg}. The "M" mass should be found between the $\eta$ and $\eta'$ masses.

If observed, the expected decay in two $\gamma$'s should be narrow, allowing to ignore possible theoretical mesons as 
$\kappa$ or $\sigma$. The predicted estimated mass and width of the (controversial) $\kappa$ are \cite{pdg} in a broad range M$_{\kappa}  \approx$ 600-900~MeV and $\Gamma_{\kappa}  \approx$ 400-770~MeV. The Breit-Wigner width of the $f_{0}$(500) ($\sigma$) meson is estimated \cite{pdg} to be $\approx$  400-700~MeV.
\section{Experimental scrutiny of known experimental data}
Having the $\eta$ quantum numbers, all possible "M"  strong decays are forbidden in lowest order (except three pions), as well as first order electromagnetic decays, except the decay to 
$\pi^{+} \pi^{-} \gamma$, commented above. The main allowed decay is the second order electromagnetic transition "M" $\to 2\gamma$. Disintegrations of other mesons ($\rho$, $\omega$, ...)  by weak interaction are allowed, but will give very small signals in the experimental spectra, and are therefore neglected. Since the "M" mass is larger than the $\eta$ mass, we have to consider three other disintegration modes: the $\eta \pi$ is forbidden by isospin and the $\eta \gamma$ by charge conjugation; the mass of $\eta\pi\pi$ is too large.   

The neutral decay modes of the $\eta$ meson were discussed in \cite{nefkens}. 
\subsection{The two photon invariant mass data}
The meson production in ${\bar p}{\it p}$ annihilation was studied by the Crystal Barrel Collaboration at LEAR (CERN) \cite{amsler} \cite{amsler2}.

\begin{center}
\begin{figure}[ht]
\includegraphics[scale=0.42] {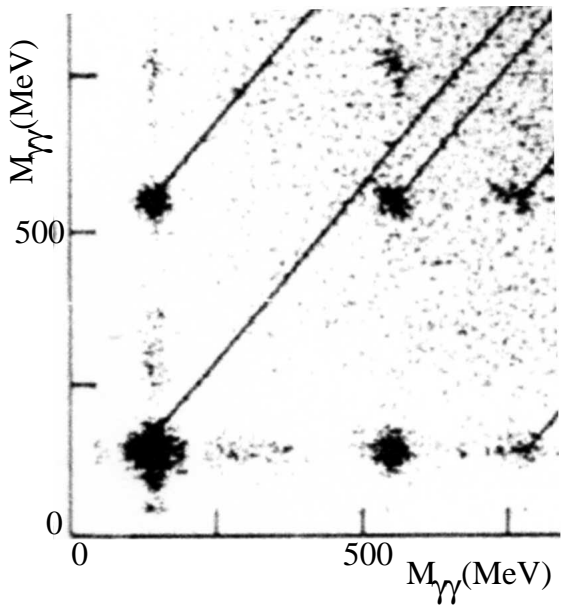}
\caption{Part of Fig.~6 \cite{amsler2}  showing the scatter plot of the two photon invariant mass $m(\gamma1, \gamma2)$ (in MeV), versus the two photon invariant mass $m(\gamma3, \gamma4)$ (in MeV). Reprinted figure with permission from Claude Amsler, Reviews of Modern Physics, Vol. 70, 1293 (1998). Copyright 2014 by the American Physical Society.}
\label{Fig:fig1}
\end{figure}
\end{center}

In the review article on the proton-antiproton annihilation results from Lear, a bi-dimensional spectra of 
2$\gamma$ invariant-mass distribution versus a 2$\gamma$ invariant-mass distribution, is shown for a sample of 4$\gamma$ events in Fig. 6 of Ref. \cite{amsler2}. The data are not available in tabulated form, and are reproduced in Fig. \ref{Fig:fig1}. The intensity of the blobs decreases with increasing masses. They are connected in the original figure by straight lines,and named by the authors  successively as  $\pi^{0}\pi^{0}$,  $\pi^{0}\eta$, $\eta\eta$, and $\pi^{0}\omega$. Since the charge conjugation forbids the $\omega$ disintegration into two pions (except by weak interaction) we considered carefully the masses of the corresponding blobs close to 750~MeV and excited with a moderate intensity.  Due to unprecise shapes of the useful blobs, the extracted masses are also unprecise.  We obtain the following approximate values (in MeV): when m(y) = $\pi^{0}$ then m(x) = 768~MeV; when m(y) = m($\eta$), then m(x) = 762; when m(x) = m($\eta$), m(y) = 753. The mean value between these three masses is M$\approx$761~MeV.

This blob close to M = 760~MeV was identified to $\pi^{0}\omega$.
It was written, in several papers from this ${\it p}{\bar p}$ annihilation at rest: "for the $\omega$ signal, a low energy photon has escaped detection". For example in the two-$\gamma$ invariant mass distribution for $\pi^{0} \gamma \gamma$ events \cite{amsler5},  the peak named $\omega$ is located close to 763~MeV. 

Also in another paper \cite{klempt}, the authors show a spectrum of the $\gamma \gamma$ momentum. Their Fig. (7b) shows, after suppression of the $\pi^{0} \pi^{0}$ signal, an asymmetric structure in the region of the $\omega$ bump, which is eliminated by introducing cuts on the $\gamma \gamma$ pair. 

 In the spectra of the invariant $\gamma\gamma$ mass for 
${\it p}{\bar p} \to \pi^{+} \pi^{-} \gamma \gamma$ events \cite{amsler1}, the authors associate the signal close to M = 760~MeV, to events where one soft photon from the decay $\omega \to \pi^{0} \gamma$ is missing. 

 However the spectra of a single $\gamma$ in the missing $\pi^{0}$ 
rest frame is given in  Fig.~13 of Ref. \cite{amsler2}. It shows that the
probability to have photons of energies E$\le$20~MeV is very
small. Moreover, in case of a loss of low energy photons, we will expect a
continuous blob built with the remaining photons and not a peak with a
width close to that of $\eta'$. Therefore the assumption advocated for the $\omega$ signal is questionable. The existence of "M" close to 762~MeV seems to solve the difficulty.

\begin{figure}[ht]
\hspace*{-0.2cm}
\scalebox{0.6}[0.5]{
\includegraphics[bb=21 269 522 517,clip,scale=0.8] {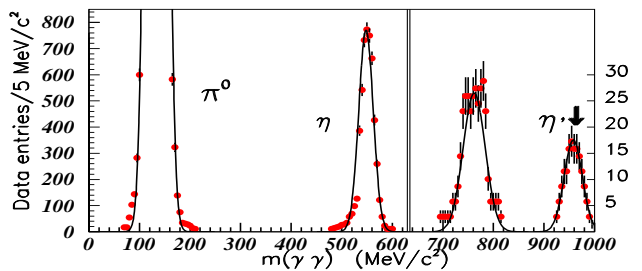}}
\caption{Color online. $\gamma \gamma$ invariant mass of the of the ${\it p}{\bar p}$ annihilation  into 
$\pi^{+} \pi^{-} \gamma \gamma$ events \cite{amsler1}. See text.}
\label{Fig:fig2}
\end{figure}

The data of Fig.~1 in Ref. \cite{amsler1}, are  read and reported in Fig.~\ref{Fig:fig2} of the present paper. 
The events under discussion are fitted with a gaussian centred at M = 762~MeV. The relative amplitudes from $\pi^{0}/\eta$/"M"/$\eta'$ are $\approx$ 490/40/1.5/1, respectively. 

\begin{figure}[ht]
\hspace*{-0.2cm}
\scalebox{0.65}[0.6]{
\includegraphics[bb=36 241  530 547,clip,scale=0.8] {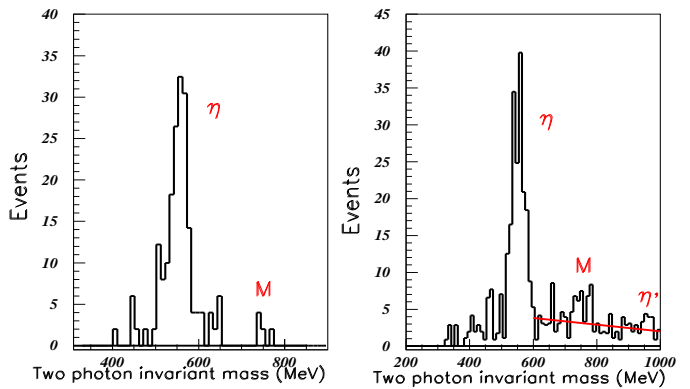}}
\caption{Color online. $\gamma \gamma$ invariant mass from ECAL barrel calibration at LHC startup with $\eta \to \gamma\gamma$ \cite{litvin}.}
\label{Fig:fig3}
\end{figure}

The invariant mass of two photons was studied during the investigation to calibrate the CMS electromagnetic calorimeter \cite{litvin}. The data are read from their Figs. 5 and 6, and reported in 
Fig. \ref{Fig:fig3} which shows, in both inserts, the two photon invariant mass after different cuts. Although the statistics is poor, an increase of the number of events is observed around $M_{\gamma\gamma}$ = 750~MeV. The events below $\eta$ mass correspond to exotic mesons weakly excited, already observed at M = 415 and 482~MeV \cite{boregle}.

The ratio between "M" and $\eta'$ is close to  the one observed in Fig.~\ref{Fig:fig2}.
\subsection{Other reactions}
 The cross section of $e^{+} e^{-} \to \eta \gamma$ reaction was studied at the VEPP-2M $e^{+} e^{-}$ collider \cite{achasov1}.
The $\eta$ was selected by the $\pi^{+} \pi^{0} \pi^{-}$ channel (open for $\omega$ disintegration) and  3$\pi^{0}$ channel  (forbidden for $\omega$ by charge conjugation). The spectra
was studied in the useful range: 755.26$\le$E$\le$1055.64~MeV, where E is the centre of mass energy. It covers the range between $\eta$(548) and $\eta$'(958). However the binning is only precise close to   
the $\eta$(548) and $\eta$'(958) masses and not at all close to E$\approx$912~MeV which will correspond to "M", preventing to get any information. The $\eta$ was also selected by the 3$\pi^{0}$ channel, forbidden for $\omega$ by charge conjugation.

The cross section of the same reaction when the $\eta$ was selected  by its 3$\pi^{0}$ decay mode was studied in a wider energy range, but  the 30~MeV binning and the  low counting rate, prevents us to use  these data.

The rare decay $\eta \to \pi^{0}\gamma\gamma$ was studied with the Crystal Ball /TAPS detectors at the Mainz Microtron \cite{nefkens3}.  The data were only given for masses lower than 680~MeV. The same experiment has been also studied with the Crystal Ball at the AGS \cite{prakhov3}. The limit set here on the two photon invariant mass spectra is 420~MeV.

 The reaction $e^{+} e^{-} \to  \pi^{+} \pi^{-} \gamma$ was studied  by the KLOE Collaboration at 
 DA$\Phi$NE \cite{alosio}. The "M" mass corresponds exactly to the large "anomaly" observed in the cross section. The
 $(\pi^{+} \pi^{-} \gamma)$ is not forbidden, but its branching ratio is small 4.6 \% for 
$\eta$(548) and decreases for increasing masses as already mentioned.
The 3$\pi^{0}$ invariant mass spectra in the $\omega$ region was studied  
with the Crystal Ball multiphoton spectrometer at the Mainz Microtron MAMI \cite{starostin}. The spectra shown in Fig.~4 (left) of Ref. \cite{starostin} is regular, without  any enhancement above a rather large background, in the "M" mass region. The authors wrote that " the main background is the $\gamma p \to 3\pi^{0} p$ direct production". It should also be noticed that the branching ratio for the disintegration  of a meson with the $\eta$ quantum numbers into 3$\pi^{0}$ decreases with the mass:  from $\eta(548)$ where it corresponds to 32.57 \% to 1.68 10$^{-3}$ for $\eta$'(958). 

The ${\bar p}{\it p}$ annihilation in flight was studied by the Crystal Barrel Collaboration at Lear \cite{abele}, into $\omega\pi^{0}$, $\omega\eta$, and $\omega\eta$' channels, therefore unable to give information concerning the present study.

Following the discussion above, we conclude that the existing data do not exclude the possible existence of the "M" meson.The mean value for its mass is taken to be M = 762~MeV.  This has to be confirmed,  and  deserves  dedicated precise measurements.

\section{Discussion}


In the systematics of Table I,  we noted the absence  of  a (${\it s}{\bar s}$) S=0 meson corresponding to the  (${\it s}{\bar s}$) S=1  $\phi$ meson.

\begin{figure}[hb]
\hspace*{-0.2cm}
\scalebox{0.6}[0.9]{
\includegraphics[bb=32 346 503 516,clip,scale=0.8] {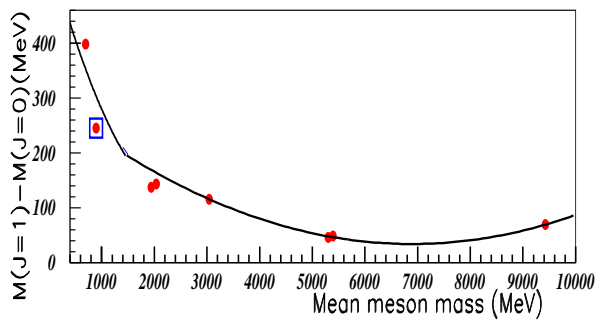}}
\caption{Color online. Mass difference between the first (S =1) minus the second (S=0) meson mass families, plotted versus the corresponding mean masses.}
\label{Fig:fig4}
\end{figure} 

Fig.~\ref{Fig:fig4} shows the mass difference between the first states S = 1 and S = 0, of 
the eight meson families, from ${\it q}{\bar s}$ to ${\it b}{\bar b}$. The differences are plotted versus the corresponding mean meson masses. A line to drive the eye is shown between the points corresponding to the known masses. The mass difference between both  (${\it s}{\bar s}$) mesons, namely the $\phi$ at 
M = 1019.5~MeV and the "M" at M = 762~MeV (surrounded by blue square) lie along this line. This curve is consistent with the data on the variation of the hyperfine splitting versus spin averaged meson multiplet mass shown in \cite{felipe}. In this last work, the experimental values, plotted versus [3*M(1$^{--}$) + M(0$^{-+}$)]/4, are close to RPA calculations.
\subsection{Possible width of the "M" meson}
\begin{figure}[h]
\hspace*{-0.2cm}
\scalebox{0.5}[0.5]{
\includegraphics[bb=4 140 517 543,clip,scale=0.8] {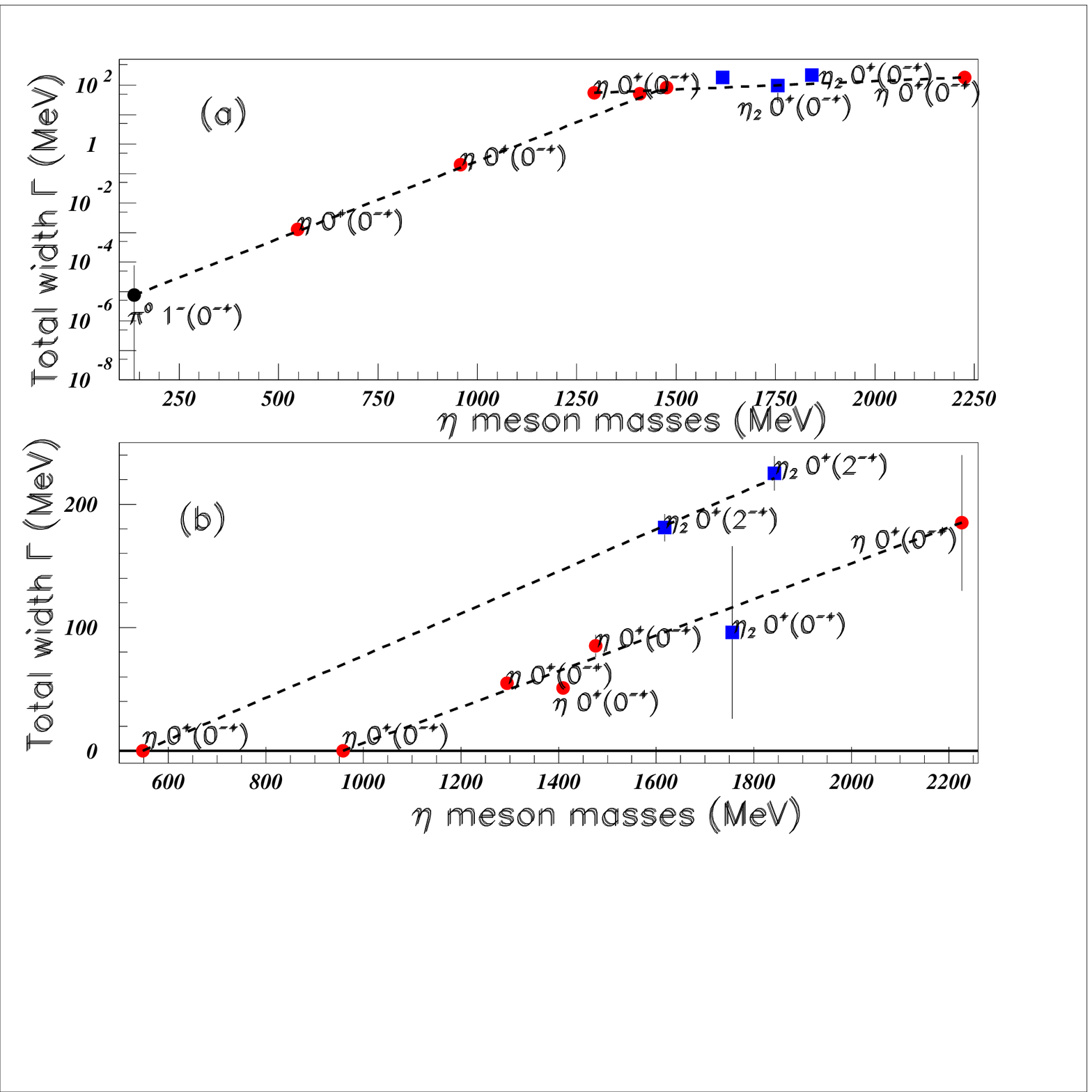}}
\caption{Color online. $\eta$ mesons total width $\Gamma$(MeV). Full red circles correspond to 
$\eta$, full blue squares to $\eta_{2}$. See text.}
\label{Fig:fig5}
\end{figure}

The width of the "M" meson, is expected to be much smaller than experimentally observed, as the data are driven by the experimental resolution. Fig.~\ref{Fig:fig5} shows in insert (a), in log scale, increasing widths of the first pseudo-scalar
$\eta$ mesons, incidentally aligned with the pion width. Insert (b) shows, in linear scale, the alignment between the $\eta$ widths for masses M$\ge$0.95 GeV.  By interpolation, we tentatively predict the total "M" width $\Gamma \approx$ 18 keV.
\subsection{Observation of mass symmetries}
\begin{figure}[h]
\hspace*{-0.2cm}
\scalebox{0.58}[0.58]{
\includegraphics[bb=16 131 525 548,clip,scale=0.8] {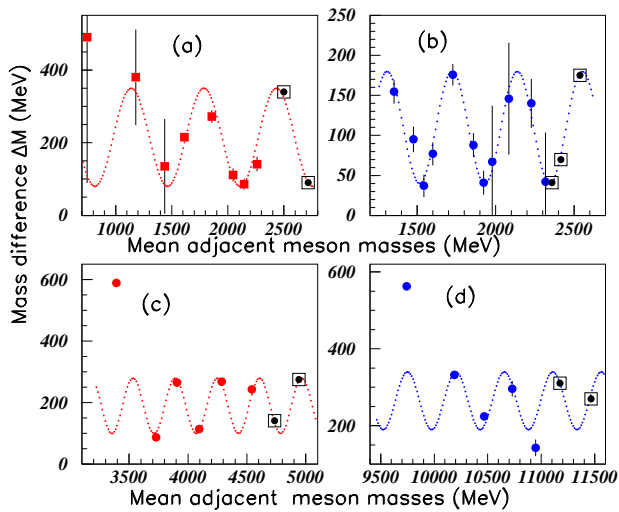}}
\caption{Color online. Mass difference between successive masses, plotted versus the same mean masses of a given meson family. Inserts (a), (b), (c), and (d) show respectively the data for $f_{0}$, $f_{2}$, ${\it c}{\bar c}$, and ${\it b}{\bar b}$ mesons. See text.}
\label{Fig:fig6}
\end{figure}

Let us point out that the meson masses display symmetries. One of such symmetries is shown below, first applied to several meson families, then applied to $\eta$ mesons. Fig.~\ref{Fig:fig6} shows the mass difference between successive masses, plotted versus the mean value of the same masses. Such plot is very convenient to emphasize regularities in the mass series. 
Such plot can be done only for families holding at least five masses.
All fits in Fig.~\ref{Fig:fig6} are obtained using a cosine function:
$$
\Delta M = \alpha_{0} + \alpha_{1}\cos[(M - M_{0}) / M_{1}],
$$
where M$_{0}$ /M$_{1}$ is defined within 2$\pi$. All coefficients, and masses used to draw the figure are in MeV. The values of the parameters are given in Table II. 

Insert (a) shows these data for the $f_{0}$ $0^{+}(0^{++})$ unflavoured mesons. The very large error bar of the first data is due to the broad and badly determined mass of the $f_{0}$(500) or $\sigma$. Extrapolating the fit allows to predict the next possible $f_{0}$ not yet observed. The masses, drawn by black full circles surrounded by black squares, are M $\approx$ 2670 and 2760~ MeV.

Insert (b) shows the data for $f_{2}$ $0^{+}(2^{++})$ unflavoured mesons. The extrapolation indicates the next possible ${\it f_{2}}$ masses: M $\approx$ 2380, 2450, and 2625~MeV.

Insert (c) shows the data for the $0^{-}(1^{--})$ charmonium ${\it c}{\bar c}$ mesons. The mass of the last quoted such meson X(4660) $?^{?}(1^{--})$ fits perfectly in this  distribution, and is therefore kept.  The extrapolation allows to predict tentatively the next corresponding masses: M $\approx$ 4805 and  $\approx$5080~MeV. The first point ($\Delta$M$\approx$590~MeV) lies outside the fit. After a possible arbitrary introduction of a not (yet ?) observed charmonium at M  $\approx$ 3385 $\pm$ 50~MeV between the two masses: M  $\approx$ 3096.916 and  $\approx$ 3686.109~MeV, we obtain two new points which replace the previous data point at $\Delta$ M = 590~MeV and lie on the distribution.
\begin{figure}[h]
\hspace*{-0.2cm}
\scalebox{0.58}[0.58]{
\includegraphics[bb=18 225 518 548,clip,scale=0.8] {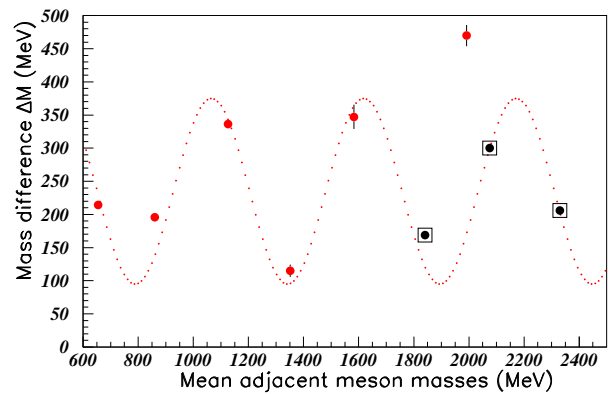}}
\caption{Color online. Mass difference between successive masses, plotted versus the same mean masses of $\eta$ mesons.  See text.}
\label{Fig:fig7}
\end{figure}

Insert (d) shows the data for the $\Upsilon$ $0^{-}(1^{--})$ bottomonium ${\it b}{\bar b}$ mesons.  Here, again, the agreement between fit and data,  is improved by the arbitrary introduction of a new mass at M $\approx$  9695 $\pm$50~MeV.  The tentatively extrapolated masses at the large mass side are: M  $\approx$11330 and  $\approx$11560~MeV, shown by black full circles surrounded by black empty squares. 

Fig.~\ref{Fig:fig7} shows similar data for $\eta$ mesons. The introduction of the M = 762~MeV $\eta$ meson mass would perfectly fits into the systematics except of the known $\eta$ masses. The extrapolation of this behaviour would suggest  a missing $\eta$ mass at M$\approx$1925~MeV and a possible next $\eta$ mass at M$\approx$2430~MeV. The resulting data are shown by black full circles enlarged by black squares in Fig.~\ref{Fig:fig7} .

None of these possibly missing masses is reported in the section "Other light mesons, further states" \cite{pdg}.
\begin{table}[h]
\caption{Coefficients (in MeV) of the fits shown in Figs.~6 and 7.} 
\label{Table VII}
\vspace{5.mm}
\begin{tabular}{c c c c c c c}
\hline
meson&qua. num.&fig.&$\alpha_{0}$&$\alpha_{1}$&$M_{1}$\\
\hline
unflav. $f_{0}$&$0^{+}(0^{++})$&6(a)&135&215&103\\
unflav. $f_{2}$&$0^{+}(2^{++})$&6(b)&70&110&66\\
charm. $c-{\bar c}$&$0^{-}(1^{--})$&6(c)&90&190&57\\
bottom. $b-{\bar b}$&$0^{-}(1^{--})$&6(d)&75&265&72\\ 
\hline
unflav. $\eta$&$0^{+}(0^{-+})$&7&140&235&88\\
\hline
\end{tabular}
\end{table}
\begin{figure}[hb]
\hspace*{-0.2cm}
\scalebox{0.9}[0.8]{
\includegraphics[bb=21 237 274 543,clip,scale=0.8] {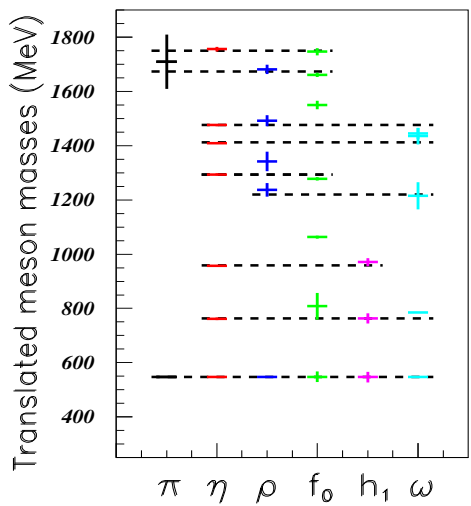}}
\caption{Color online. Low masses (in MeV) of several unflavoured meson families. See text.}
\label{Fig:fig8}
\end{figure}

Another symmetry property is observed when looking at the shifted masses of several unflavoured meson families. A translation is applied to the masses, in order to equal the first mass of each series to the value of first $\eta$ meson. The translation amount is: 410~MeV for $\pi$, 0 for $\eta$, -228~MeV for $\rho$, -442~MeV for f$_{0}$, -623~MeV for h$_{1}$, and -235~MeV for $\omega$. We observe, after translation, the mass correspondence of several unflavoured meson families, Fig.~\ref{Fig:fig8}. For example, the translated second excited masses of $\eta$, "M"(762), $h_{1}$, $f_{0}$ inside error bar, and $\omega$ are equal. The $\phi$(1020) and $\phi$(1680) are reported together with the $\omega$ data, since they have the same quantum numbers. 

Since the quantum numbers are different for the different families, it is suggested that the quantum numbers define the global mass scale instead of the mass differences within a family. 

From Fig. \ref{Fig:fig8}  it appears that, in order to fulfil completely the systematics, several not yet observed  mesons may exist.

\begin{figure}
\hspace*{-0.2cm}
\scalebox{0.9}[0.5]{
\includegraphics[bb=24 230 280 550,clip,scale=0.8] {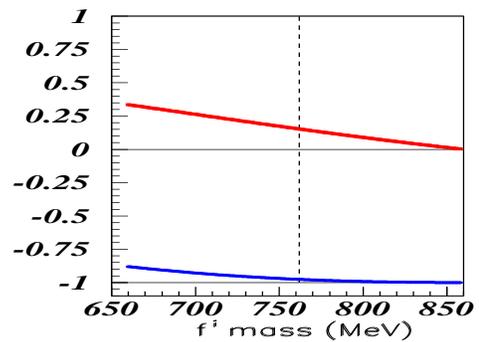}}
\caption{Color on line. Variation versus the "M" mass (in MeV) of the (${\it u}{\bar u}$ + ${\it d}{\bar d}$) amplitude (red squares)  and of the ${\it s}{\bar s}$ amplitude (blue circles). }
\label{Fig:fig9}
\end{figure} 
 \subsection{Mixing between pseudoscalar mesons} 
The following discussion is directly inspired from the "Quark Model" chapter \cite{amsler3} developed in PDG \cite{amsler3}. The suggestion is made that the four light mesons, building the middle place of the pseudoscalar multiplet, are the $\pi^{o}$, $\eta$, "M", and $\eta_{c}$.
The $\eta$'(958) joins then the several light unflavoured mesons which cannot find room in the $q\bar q$ quark model assignement shown in \cite{amsler3} table 14.2.
Among these mesons, we have today three $^{1}S_{0}$ (including $\eta$'(958)), six $^{3}P_{0}$, ten $^{3}P_{2}$, and a reduced number of other $J^{PC}$ mesons. Of course, some of these mesons might consist of four quarks, or glueball states, and therefore be exotic.

 It was shown in \cite{amsler3} that the vector meson masses induce a vector
 mixing angle $\theta_{V}$ = 35$^{0}$ between the SU(3) wave functions $\Psi_{8}$ and $\Psi_{1}$, allowing a very close to ideal mixing, the $\phi$(1020) being a nearly pure 
(${\it s}{\bar s}$). This was not the case for pseudoscalar mesons.

The same calculation, as the one shown in \cite{amsler3} is done using the following masses (and the same notations): m(K$^{0}$) = 497.614~MeV, m(a) = m($\pi^{0}$) = 134.977~MeV, m$_{f}$ = m($\eta$) = 547.853~MeV, and  m$_{f}$' = "M" = 762~MeV. The resulting 
pseudoscalar mixing angle $\theta_{PS}$ = 22.77$^{o}$ as extracted from equation  (14.9) of \cite{amsler3}. It results a quasi neat ($s\bar s$) for the "M":
\begin{equation}
M=0.153(u\bar u + d\bar d) - 0.976( s\bar s)
\end{equation}
and a mixed  ($u\bar u + d\bar d$)  and ($s\bar s$) for the $\eta$(547.85):
\begin{equation}
m(548) = 0.690 (u\bar u + d\bar d) + 0.216 ( s\bar s)
\end{equation}

Fig.~\ref{Fig:fig9} shows the variation of both amplitudes (${\it u}{\bar u}$) + ${\it d}{\bar d}$)  (full red squares) and ${\it s}{\bar s}$ (full blue circles) versus the variation of the f' mass.
When using charged masses $\pi^{\pm}$ instead of $\pi^{0}$ and $K^{\pm}$ instead of $K^{0}$, the variations are almost the same.

\begin{figure}[hb]
\hspace*{-0.2cm}
\scalebox{0.95}[0.5]{
\includegraphics[bb=27 239 289 552,clip,scale=0.8] {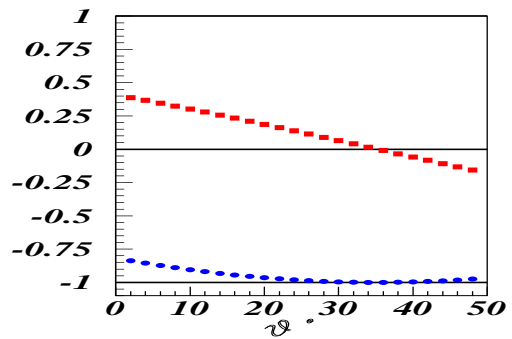}}
\caption{Color online. Variation versus the mixing angle $\theta$ of the (${\it u}{\bar u}$ + ${\it d}{\bar d}$) amplitude (red squares) on one side, and of the ${\it s}{\bar s}$ amplitude (blue circles) on the other side, obtained using the $\pi^{0}$ and $K^{0}$ meson masses.}
\label{Fig:fig10}
\end{figure} 

Fig.~\ref{Fig:fig10} shows the variation of both amplitudes (${\it u}{\bar u}$) + ${\it d}{\bar d}$)  (full red squares) and ${\it s}{\bar s}$ (full blue circles) versus the variation of the mixing angle $\theta$. The variation of the ${\it s}{\bar s}$ amplitude is rather slow around $\theta$ = 30$^{0}$ (close to -1 from $\theta$ = 20$^{0}$ to $\theta$ = 50$^{0}$, when the  (${\it u}{\bar u}$) + ${\it d}{\bar d}$) amplitude varies in the same interval from 0.21 up to -0.06. The limits on $f^{'}$ mass shown in Fig.~~\ref{Fig:fig9},  corresponds to the angular limits:  3.6$^{0}$ and 35.5$^{0}$ on Fig.~~\ref{Fig:fig10}.
\section{Conclusion}
A careful analysis of already known data, allows us to predict the existence of a low mass 
pseudoscalar S = 0 "M" meson, at M $\approx$ 762~MeV, having an important ${\it s}{\bar s}$ amplitude in its wave function, and therefore being the spin = 0 counterpart of the vector meson $\phi$.  Its existence allows to solve the difficulty recalled in the introduction concerning the mixtures of the SU(3) wave functions for the pseudoscalar mesons.

This quasi pure amplitude (${\it s}{\bar s}$) for  "M",  gives a possible explanation to the absence of the observation of S=0 low mass pseudoscalar meson in existing data. Its mass is close to the $\omega$ mass, and it is too light to disintegrate into two kaons. The relatively small $\gamma \gamma$ invariant mass was attributed to a remnant of $\omega$ disintegration. 
This assumption was already commented and criticised. We note also that the $\omega$ decay mode into 3$\gamma$'s is very small, as small as the charge conjugation violating mode. So the moderate intensity of the signal is likely due to the relatively small production of the 
$ qqq-\bar q\bar q\bar q \to  s\bar s$ reaction.

We are very grateful to Professors Claude Amsler and Alex Bondar for their interest and helpful remarks.
 

\begin{thebibliography}{99}
\bibitem{godfrey}S. Godfrey and N. Isgur, Phys. Rev. D {\bf 32}, 189 (1985).
\bibitem{amsler3}C. Amsler, T. DeGrand, and B. Krusche, Phys. Rev. {\bf D86}, 199 (2012).
\bibitem{hooft}G. t'Hooft, Phys. Rept. {\bf 142}, 357 (1986).
\bibitem{close}F. Close, S. Donnachie, and G. Shaw, in Electromagnetic Interactions and Hadronic Structure, Cambridge Monographs on Particle Physics, Nuclear Physics, and Cosmology, Cambridge University Press (2009).
\bibitem{pdg}J. Beringer {\it et al.} (Particle Data Group), Phys. Rev. D {\bf 86}, 010001 (2012).
\bibitem{nefkens}B.M.K. Nefkens and J.W. Price, arXiv:nucl-ex/0202008v1 (2002).
\bibitem{amsler}C. Amsler {\it et al.}, Phys. Lett. B {\bf 327}, 425 (1994).
\bibitem{amsler2}C. Amsler, Reviews of Modern Physics, {\bf 70}, 1293 (1998).
\bibitem{amsler5}C. Amsler, AIP Conference Proceedings {\bf 243},
 263 (1992).
\bibitem{klempt}C. Amsler {\it et al.} Crystal Barrel Collaboration, Z. Phys. C{\bf 58}, 175 (1993);
E. Klempt, Acta Physica Polonica B{\bf 24}, 1813 (1993).
\bibitem{amsler1}C. Amsler {\it et al.} Crystal Barrel Collaboration, Phys. Lett. B {\bf 346}, 203 (1995).
\bibitem{litvin} V.Litvin, H. Newman, and S. Shevchenko, CMS Internal Note, CMS IN 2004/XXX.
\bibitem{boregle}B. Tatischeff and E. Tomasi-Gustafsson, Bentham Open Physics Journal, I, 15 (2014); 
POS(Baldin ISHEPP XXII) 111 (2015); arXiv:1503.0347v1 [nucl-ex] 92015). 
\bibitem{achasov1}M.N. Achasov {\it et al.}  Phys. Rev. D {\bf 74}, 014016 (2006).
\bibitem{nefkens3}B.M.K. Nefkens {\it et al}, Phys. Rev. C {\bf 90}, 025206 (2014).
\bibitem{prakhov3} S. Prakhov  {\it et al}, Phys. Rev. C {\bf 78}, 015206 (2008).
\bibitem{prakhov}S. Prakhov, Phys. Rev. Lett. {\bf 84}, 4802 (2000).
\bibitem{alosio}A. Alosio {\it et al.}, Phys. Lett. B {\bf 606}, 12 (2005).
\bibitem{starostin}A. Starostin {\it et al.} Phys. Rev. C {\bf 79}, 065201 (2009).
\bibitem{abele}A. Abele {\it et al.} Crystal Barrel Collaboration, Eur. Phys. J. C 12, 429 (2000).
\bibitem{felipe}F. J.  Llanes-Estrada, S. R. Cotanch, A. P. Szczepaniak, and E. S. Swanson, Phys. Rev. C 
{\bf 70}, 035202 (2004).

\end{thebibliography}
\end{document}